\documentclass[]{spie}  

\usepackage{amsmath,amsfonts,amssymb}
\usepackage{graphicx}
\usepackage[colorlinks=true, allcolors=blue]{hyperref}
\usepackage{afterpage,placeins}

\title{Nowcasting the turbulence at the Paranal Observatory}

\author[a]{J. Milli}
\author[b,c]{R. Gonzalez}
\author[b]{P. R. Fluxa}
\author[a]{A. Chacon}
\author[a]{J. Navarette}
\author[d]{M. Sarazin}
\author[a]{E. Pe\~{n}a}
\author[e]{R. Carrasco-Davis}
\author[a]{A. Solarz}
\author[a]{J. Smoker}
\author[a]{C. Martayan}
\author[a]{C. Melo}
\author[a]{E. Sedaghati}
\author[a]{S. Mieske}
\author[d]{O. Hainaut}
\author[d]{L. Tacconi-Garman}

\affil[a]{European Southern Observatory, Alonso de C\'ordova 3107, Casilla 19001, Santiago, Chile}
\affil[b]{Centro I+D MetricArts, Santiago, Chile}
\affil[c]{Centro de Astro-Ingenier\'ia, Pontificia Universidad Cat\'olica, Santiago, Chile} 
\affil[d]{European Southern Observatory, Karl-Schwarzschild-Stra{\ss}e 2, 85748 Garching, Germany}
\affil[e]{Dept. of Electrical Engineering, Universidad de Chile, Santiago, Chile}

\authorinfo{Further author information: Julien Milli: E-mail: julien.milli@mines-paris.org}

\begin{document} 
\maketitle

\begin{abstract}

At Paranal Observatory, the least predictable parameter affecting the short-term scheduling of astronomical observations is the optical turbulence, especially the seeing, coherence time and ground layer fraction. These are critical variables driving the performance of the instruments of the Very Large Telescope (VLT), especially those fed with adaptive optics systems. Currently, the night astronomer does not have a predictive tool to support him/her in decision-making at night.  As most service-mode observations at the VLT last less than two hours, it is critical to be able to predict what will happen in this time frame, to avoid time losses due to sudden changes in the turbulence conditions, and also to enable more aggressive scheduling. We therefore investigate here the possibility to forecast the turbulence conditions over the next two hours. We call this "turbulence nowcasting", analogously with weather nowcasting, a term already used in meteorology coming from the contraction of "now" and "forecasting". We present here the results of a study based on historical data of the Paranal Astronomical Site Monitoring combined with ancillary data, in a machine learning framework.  We show the strengths and shortcomings of such an approach, and present some  perspectives in the context of the Extremely Large Telescope.

\end{abstract}

\keywords{Turbulence, Nowcasting, Seeing, Coherence time, Ground-layer fraction, Site monitoring}

\section{Introduction}
\label{sec_intro}

\subsection{Why nowcasting the turbulence?}
\label{sec_why}
A nowcast represents an extrapolation of the current conditions to the very near future. There are several reasons why the Paranal Observatory is interested to know how the turbulence will evolve, especially on a one- to two-hour timescale, and we summarise them below.

\subsubsection{Supporting the night astronomer in decision-making at night}

Most of the observations at the VLT are carried out in Service Mode, whose demand peaked at 87\% in the Period 100 (September 2017 - March 2018)\cite{Patat2017}. As a result, the efficiency of the observatory to execute the scientific programs present in the service queue depends upon the night astronomer, who must select the correct observation compatible with the current conditions. While this holds true for any particular conditions, including for instance the meteorological parameters that affect the operations of the VLT or the observability conditions of the target, this can be particularly challenging for the turbulence conditions. There currently exists no forecast system of the optical turbulence (hereafter OT) used in operations to support the night astronomer. The seeing, which represents the integrated strength of the turbulence along the line of sight, is the parameter used systematically for any instrument at the VLT to constrain the turbulence conditions, and this quantity can evolve rapidly in a matter of minutes. In addition, the use of adaptive optics (AO) systems at the VLT has expanded significantly in the past 3 years: all VLTI instruments are now AO-fed\cite{Woillez2019,Woillez2016} and Unit Telescope 4 is a fully adaptive telescope \cite{Madec2018}. In terms of operations, this means that new turbulence parameters have to be considered to ensure that these systems deliver their full performance. We therefore introduced in Period 101 (April 2018) the use of the coherence time for VLT extreme AO instrument SPHERE \cite{Milli2017_SPHERE}, in P102 (September 2018) we started using the fraction of turbulence in the ground layer (hereafter GLF) internally for the operations of the ground-layer AO systems GRAAL and GALACSI \cite{Madec2018}, and in P105 (starting in April 2020), the way turbulence conditions are constrained for Service Mode observation were completely revised to allow a uniform treatment between seeing-limited instruments and AO-fed instruments. These new parameters imply that decision-making at night becomes a complex multi-parameter problem, where support from automated tools is highly desirable. 

\subsubsection{Decreasing the time spent in out-of-constraints observations due to the turbulence} 

   \begin{figure} [h]
   \begin{center}
   \includegraphics[width=\hsize]{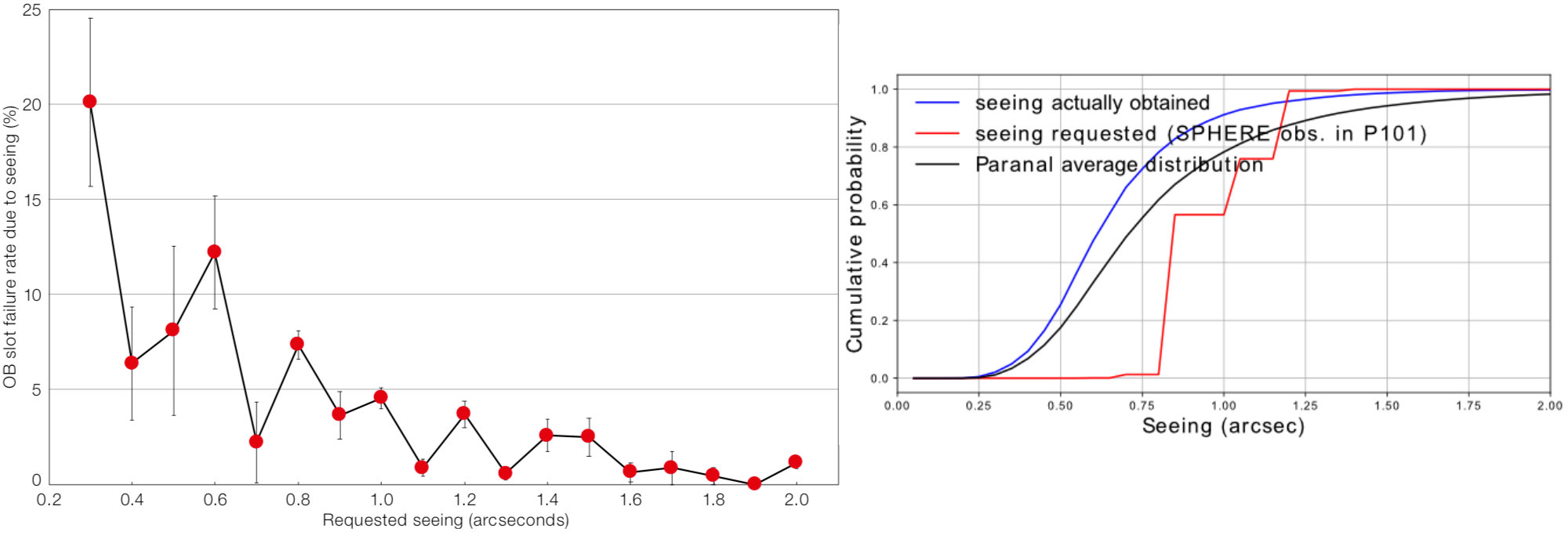}
	\end{center}
   \caption[ex] 
   {  \label{fig_failure_rate_seeing}
 Left: OB failure rate due to seeing as a function of requested seeing\cite{Rejkuba2018}. Right: Cumulative probability distribution of the seeing requested in phase 2 (red curve) compared to the delivered seeing (blue curve) for observations done with VLT/SPHERE at UT3 during Period P101. The black curve shows the distribution of the hourly-averaged seeing at the Paranal site, as measured by the Paranal DIMM from 2016 onwards. This comparison is relevant for SPHERE because most observations last about 1h, and in P101 the DIMM seeing was still used as a constraint for observations (the coherence time was introduced in P102) despite all other instruments using the image quality rather than the seeing. 
   }
   \end{figure} 

The amount of time spent on executing observations (OBs) obtained during conditions outside of the specified constraints and that are considered for repetition varies from instrument to instrument. At the VLT, on UT1 and UT2 the fraction of observations executed out of the constraints is between 5 and 10\%, while on UT3 and UT4 it is typically between 10 and 15\% \cite{Rejkuba2018}. Looking specifically at the observations declared out-of constraint due to the seeing, an independent analysis done by inspecting manually the night log of UT3 during the Period P101 (April 2018 to October 2018) where SPHERE and VISIR were the only instruments operated in Service Mode, revealed a failure rate of 4.0\%. The ESO document Cou-1628 (2015) on the policy regarding GTO\footnote{\href{https://www.eso.org/sci/observing/policies/Cou_1628_VLT_GTO_Policy_111115.pdf}{www.eso.org/sci/observing/policies/Cou\_1628\_VLT\_GTO\_Policy\_111115.pdf} } states that a night is charged 83 kEUR (in 2015 Euros) to an instrument consortium. In P101 at UT3, about 60\% of the time was dedicated to Service Mode observations. This makes a financial loss due to the seeing of 728 kEUR per year. Such a failure rate is in agreement with an independent study using UT1 to UT4 and VLTI observations carried out between Periods 90 (October 2012) and 97 (September 2016), whose core result was presented in Rejkuba et al. 2018\cite{Rejkuba2018}, that derived a $4.5\% \pm 0.6\%$ failure rate due to the seeing only. This average value does however hide an important fact: the more demanding observations are the most affected by failures due to the seeing, as shown in Fig. \ref{fig_failure_rate_seeing} (left). Therefore decreasing the time spent in out-of-constraints observations would benefit the most demanding programs to better exploit the excellent conditions of the Paranal site. This has great potential as the best conditions are mostly under-exploited, with very few proposals asking for demanding conditions, as this shown in Fig. \ref{fig_failure_rate_seeing} (right) for the example of the SPHERE instrument in Period P101.
 
\subsubsection{Enable more aggressive short-term scheduling with well-estimated risks}

Despite the fact that a failure rate of 4\% has a significant financial impact, as shown above, it still represents a small value. Indeed, looking at the pure statistics of the seeing, one would expect a failure rate of 34\% (respectively 27\%) for a seeing constraint of 0.8" (resp. 1.0") assuming that a one-hour observation is systematically started after the seeing has been below 0.8" (resp. 1") for 30min (ESO-internal study). The low achieved failure rate due to unmet seeing constraint stems likely both from a conservative choice of observation to undertake at any given opportunity by the night astronomer and from conservatism by the community, not requesting very demanding conditions (Fig. \ref{fig_failure_rate_seeing} right). This means that there is room for improvement to schedule demanding programs more aggressively if one-hour nowcast is available to help the night astronomer assess the risk undertaken by doing so. By doing so, the community could also be less conservative and dare to ask for demanding programs exploiting the best conditions of the Paranal site. 

\subsubsection{Prepare the mode of operation of the European Large Telescope (ELT)}
The ELT, located at Cerro Armazones 22\,km away from Cerro Paranal, will be an entirely adaptive telescope \cite{Bonnet2018}, feeding instruments using various flavours of AO systems, from ground-layer AO to tomographic and multi-conjugate AO systems, where the knowledge of the turbulence will be even more paramount than at the VLT to best exploit the diffraction limit of this 39m-telescope. The pressure to get telescope time will be higher than at the VLT, therefore turbulence nowcast will be even more valuable to avoid wasting telescope time and make sure the systems are optimised to deliver the best performance over the course of the observations. The VLT represents an ideal platform for a pilot study at nowcasting the turbulence. Such a pilot study will be useful to make the best choices in terms of instrumentation for turbulence monitoring at the ELT, and also to gain experience on the use of nowcast to maximise the science return of the VLT


ESO launched a call for tender to obtain a service of forecast for meteorological and turbulence parameters at Paranal observatory. However, little is known on the impact such a service could have for the efficiency of the observatory. Two studies exist on this topic (see section 1.2). The first one reports 22.8\,nights/semester spent in out-of constraints observation (without breaking down the cause, which might not be 100\% related to the astro-climate constraints, for instance airmass, moon, twilight constraints, laser collisions, missing calibration...). This translates into a failure rate of about 19\%, assuming 65\% of service mode and it gives an upper limit on the time that could be saved per semester with an accurate forecast service. The second one reports on average $4.5\% \pm 0.6\%$ failure rate due to the seeing. This gives a lower limit on the time that could be saved per semester, but it does not include the OBs aborted before completion and failed OBs because of astro-climate parameters other than the seeing. 

\subsection{Existing forecasting systems}

Although this work represents to our best knowledge the first attempt at nowcasting the turbulence using a purely empirical approach, forecasting the OT with numerical weather predictions (NWP) over longer timescales (up to several nights) has already been done in the context of astronomical observations.  

The  Advanced LBT Turbulence and Atmosphere (ALTA) system\cite{Veilleit2016,Turchi2017}, operated at the Large Binocular Telescope (LBT), is an example of such a system, currently in operations\footnote{\href{http://alta.arcetri.astro.it}{http://alta.arcetri.astro.it}} at Mount Graham International Observatory. It uses a non-hydrostatic mesoscale model (Meso-NH\cite{Lafore1998}) and a package specifically conceived for the OT and developed by INAF-OAA (Astro-Meso-NH\cite{Masciadri1999}). Short-term nowcasts are also available, by combining in situ measurements and the most recent model forecasts. The same mesoscale model was also applied to the VLT site at Cerro Paranal in a pilot study called MOSE\cite{Masciadri2013,Lascaux2013,Lascaux2015,Turchi2019}. This Paranal mesoscale model has significantly evolved since its development to increase the accuracy of the predicted parameters\cite{Masciadri2017}. A forecast system for turbulence and meteorology is also currently in operation at the Mauna Kea Observatories and is run by the Mauna Weather Center \footnote{\href{http://mkwc.ifa.hawaii.edu}{http://mkwc.ifa.hawaii.edu}}.  An alternative approach to mesoscale modelling consists in using general circulation models, although it is far less accurate. This approach was applied to Cerro Paranal to predict integrated quantities such as the seeing, coherence time or ground-layer fraction\cite{Osborn2018_GSM}. General circulation models use a much larger spatial mesh of several tens of km horizontally compared to several hundred meters for mesoscale modelling which also enable more complex boundary layer parametrization.

Our goal now is to investigate whether a simple empirical approach, based on measurements delivered from the Paranal Astronomical Site Monitor (ASM), the suite of atmospheric and turbulence sensors that equip the Paranal Observatory, can already provide nowcasts reliable enough to be used in operations.

\section{Method and results}

\subsection{Overview of the data flow}

\begin{figure} [h]
   \begin{center}
   \includegraphics[width=0.8\hsize]{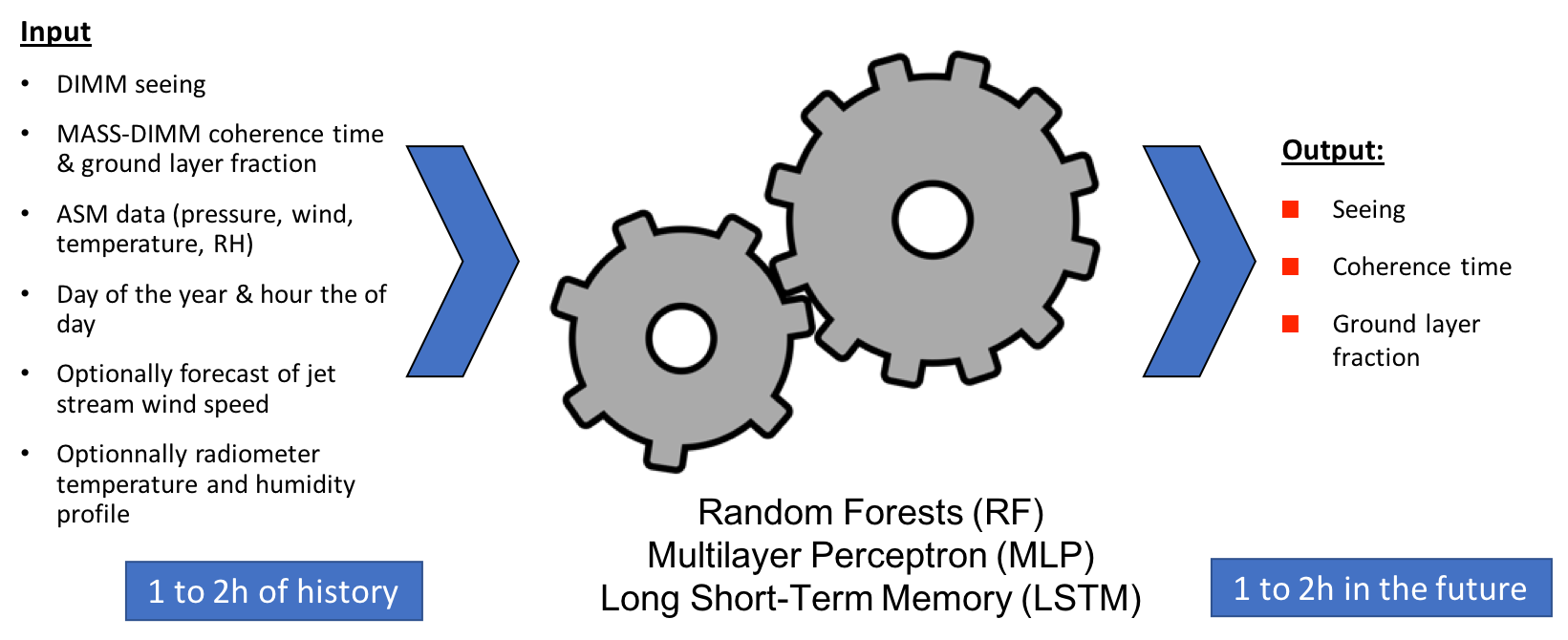}
	\end{center}
   \caption[ex] 
   {  Schematics illustrating the input/output data used, along with the types of algorithm. \label{fig_schematics_approach} 
   }
   \end{figure}

Fig. \ref{fig_schematics_approach} shows an overview of the methodology, detailing the data used as input, the expected output of the algorithm and the types of algorithms employed. 

ESO obtained the support of Microsoft and its Chilean counterpart \href{http://www.metricarts.com}{Metric Arts} in the development of the framework described in Fig. \ref{fig_schematics_approach}. 
The goal of this partnership was to show how the use of artificial intelligence could benefit the operations of the Paranal Observatory. Two proofs of concepts were selected for this purpose: the turbulence nowcast project described here, and the automatic detection of anomalies in VLT/UVES images.

The input data consists of atmospheric and turbulence data delivered by the Paranal ASM. The ASM was upgraded in April 2016 in order to solve some obsolescence issues and to support new AO requirements. We therefore used only data from April 2016 to April 2019 in this study, to have a uniform set of data. It comprises:
\begin{itemize}
\item DIMM\cite{Sarazin1990} seeing measurements with a median sampling of one point every 79s.
\item MASS-DIMM\cite{Kornilov2007} coherence time and ground-layer fraction, with a median sampling of one point every 79s.  The ground-layer fraction used here is the fraction of the turbulence in the ground layer, defined as the DIMM minus MASS seeing
\item Meteorological parameters from the Paranal Vaisala meteo tower, with a median sampling of one point every 60s. We kept here from all available sensor measurements: the air pressure (1min-average air pressure at the level of the platform), the air temperature (1min-average air temperature at 30m above the platform), and the wind speed and direction (1min-average of the U,V and W components of the wind speed measured at 20m above the platform).
\end{itemize}

Those different input data were resampled to have one data point every 5min and synchronised together. Missing data were not interpolated but left as NaN (not a number) if no data were present within the 5min sampling time. The same happened for time stamps obtained during day time, where no DIMM or MASS-DIMM data is available. As the turbulence is expected to be subject to seasonal and diurnal variations, we added at each time step the day of the year and hour of the day. We used here a cyclical representation of those two variables (cos(day/365), sin(day/365), cos(hour/24), sin(hour/24)), hence 4 more variables to capture those yearly and daily cycles. 

Those variables are either local measurements of the atmosphere at the level of the Paranal platform or integrated measurements at zenith. We however have no measurements of the meteorology in altitude. A radiometer (LHATPRO\cite{Kerber2012} for Low Humidity And Temperature PROfiling microwave radiometer) has been installed in Paranal to measure, among other, the Precipitable Water Vapour, and also provides temperature and humidity profiles up to 10km in altitude, with a variable vertical resolution from 10m (at the ground) to 1000m (at 10km) and an average sampling of one point every minute. To avoid having too many input data, we did not include those measurements as input data for the analysis, and leave this for a follow-up study. However, we kept the option of using data from the European Centre for Medium range Weather Forecast (ECMWF). Out of the available variables provided by the ensemble sampler of the ECMWF operational data (all meteorological parameters at 25 pressure levels above Paranal), we selected the 200mbar wind speed, in order to still have an idea of the jet stream wind velocity. This data is however provided with a different sampling of one point every hour. The reason why this parameter might have an impact is that the coherence time is given by $\tau_0=0.315\frac{r_0}{v_0}$\cite{Roddier1981} where $r_0$ is the Fried parameter and $v_0$ is the horizontal turbulence velocity\cite{Masciadri2001} and it was shown\cite{Sarazin2002} that $v_0$ can be well approximated by $v_0=\text{max}(v_\text{ground},0.4v_{200\text{mbar}})$ where $v_\text{ground}$ is the wind speed measured at 30m above the platform and $v_{200\text{mbar}}$ is the predicted wind speed at 200mbar.

As output variables, we are interested to predict the DIMM seeing, the MASS-DIMM coherence time and the MASS-DIMM ground layer fraction, that are all three used in real time for scheduling of VLT observations. This means the problem we are trying to solve is a regression task: we want to predict a numerical value given some input. We decided to focus on three different machine-learning algorithms. First we selected the random forests (RF) regression\cite{Breiman1984} as a baseline scenario for its ease of implementation and few model parameters. Then we tested a fully connected artificial neural network, or multilayer perceptron (MLP), and last we tried a probabilistic approach based on a recurrent neural network known as the Long Short Term Memory (LSTM).

\subsection{Data preparation}

A required preliminary step consists in preparing the data for the regression task. Because we are dealing with time series, we have to prepare the input/output data in blocks and split those blocks between a training set and a validation set. 
Each block consists of a 2h historical time series of the input variables and the corresponding 2h future times series of the desired output variable called here Y (the seeing, coherence time or GLF). This means that considering our baseline 11 input variables (Y, pressure, temperature, wind U, wind V, wind W, with a 5min sampling, the jet stream wind speed with a 1h sampling, cos(day/365), sin(day/365), cos(hour/24), sin(hour/24) with only one occurence), each block consists of a vector of $6\times24+1\times3+4\times1=151$ values as input and a vector of 24 values as output. 

Then we built all possible blocks from the 4 years of data collected. The validation set consists of data obtained before October, 1 2018 while the training set consists of data obtained after this date. This resulted in 35663 blocks for training and 5251 blocks for validation, or 87\% of the data used for training. We highlight that two different blocks can be partially overlapping. 

\subsection{Random Forests (RF)}

\begin{figure} [h]
   \begin{center}
   \includegraphics[width=1.05\hsize]{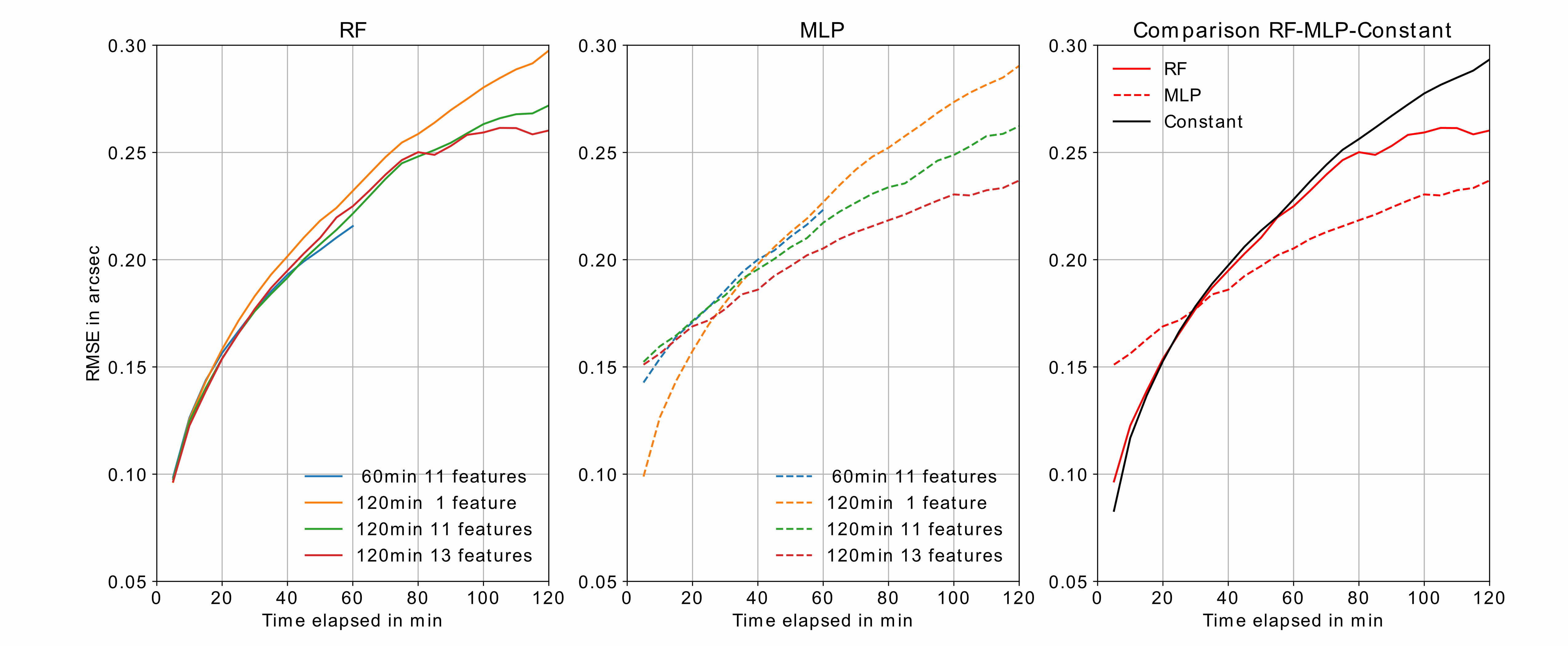}
	\end{center}
   \caption[ex] 
   {Left: error in the prediction of the seeing using random forest (RF) as a function of the elapsed time since prediction. The baseline scenario (green curve) uses 11 input variables to predict the seeing over 2h. The red curve uses 2 additional input parameters, the coherence time and the GLF to predict the seeing over 2h. The orange curve uses a single input parameter, the seeing, to predict the seeing over 2h and the blue curve is identical to the baseline scenario but only tries to predict the seeing over 1h. Middle: Same plot for the multilayer perceptron instead of the RF. Right: Comparison between RF and MLP for the best setup (120min 13 features). We overplotted the error obtained in a scenario called "Constant" where we assumed the seeing is constant over 2h, and equal to the average of the previous 15min. \label{fig_RMSE_RF} 
   }
   \end{figure}

We used the random forest regressor implemented in the Python package Scikit-learn\cite{scikit-learn}. We initialised the algorithm with 3000 trees, and used the mean squared error to measure the quality of the regression. Fig. \ref{fig_RMSE_RF} (left) shows the error in the prediction for four different setups, as estimated from the validation data set. 

Fig. \ref{fig_example_RF} shows a few examples of 2h nowcasts in the baseline scenario using RF and drawn from nights in the validation set, plotted against the seeing as measured by the DIMM.


\subsection{Multilayer perceptron}

We built a standard feedforward neural network, with a regular fully-connected (dense) network layer through the Keras\cite{Keras} Python implementation. Each layer has 32 nodes and uses the rectifier linear unit (RELU) activation function. We used the Adam gradient-based optimiser\cite{Kingma2014}, and the mean squared error as the loss function. We added as many layers as we had input variables, e.g. 11 in the case of our baseline scenario. The error in the prediction is shown in Fig. \ref{fig_RMSE_RF} (middle), it was estimated from the validation data set.

%

\subsection{Results}

Independently of the algorithm and the input variables tested, the accuracy in terms of root-mean-square error (RMSE) is between 0.2" and 0.25" for a prediction at 1h. The difference between algorithms is visible between 1h and 2h, where the MLP performs better and manage to maintain the accuracy below 0.24" whereas the RF accuracy is worse than 0.26". 
Using only a single input variable, the accuracy of RF and MLP is almost identical. Considering more input variables, MLP performs better for a prediction after 30min, while RF performs better for a prediction below 30min. 

  \begin{figure} [h]
   \begin{center}
   \includegraphics[width=1.0\hsize]{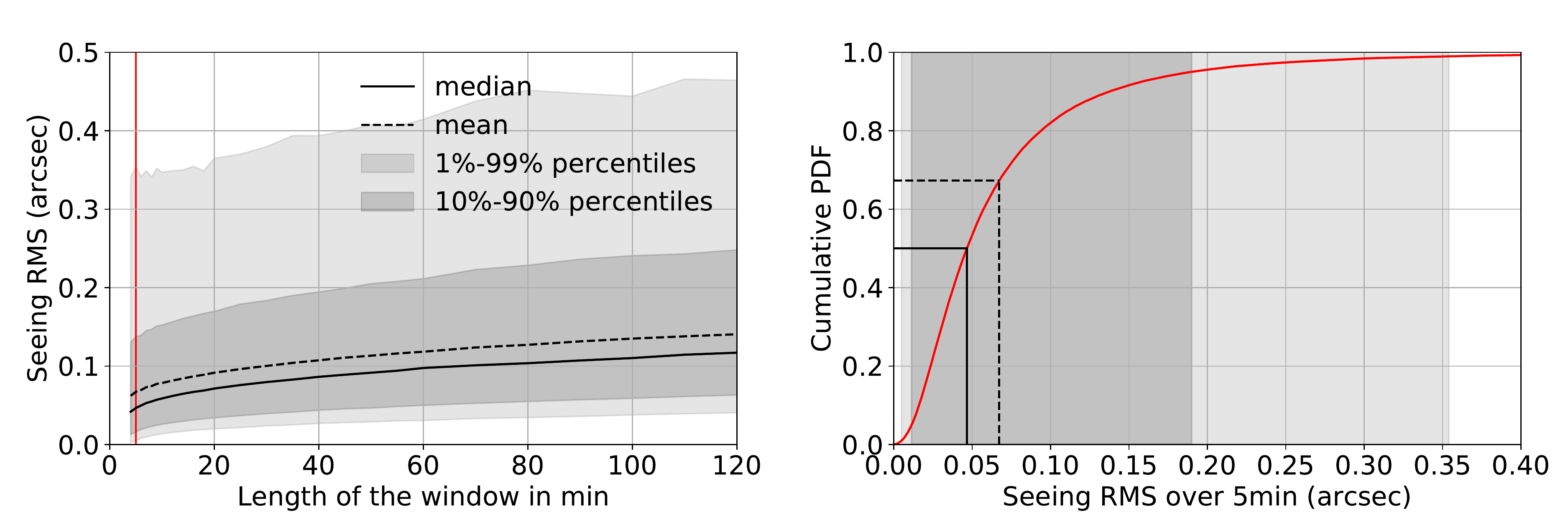}
	\end{center}
   \caption[ex] 
   { 
 Left: Intrinsic variability of the seeing, measured as the seeing RMS in a given time window, as a function of the length of the window. The data considered for this analysis are all DIMM data collected between April 2016 and April 2019 in their original 1 to 2\,min data sampling. For each window length, the graph shows the mean, median as well as the 1\%, 10\%, 90\% and 99\% percentiles to highlight the large tail in the distribution. For a 5min window (red vertical line on the left graph), the right graph shows the full cumulative distribution of the RMS.
   \label{fig_seeing_variability} }
   \end{figure} 

\begin{figure} [h]
   \begin{center}
   \includegraphics[width=0.85\hsize]{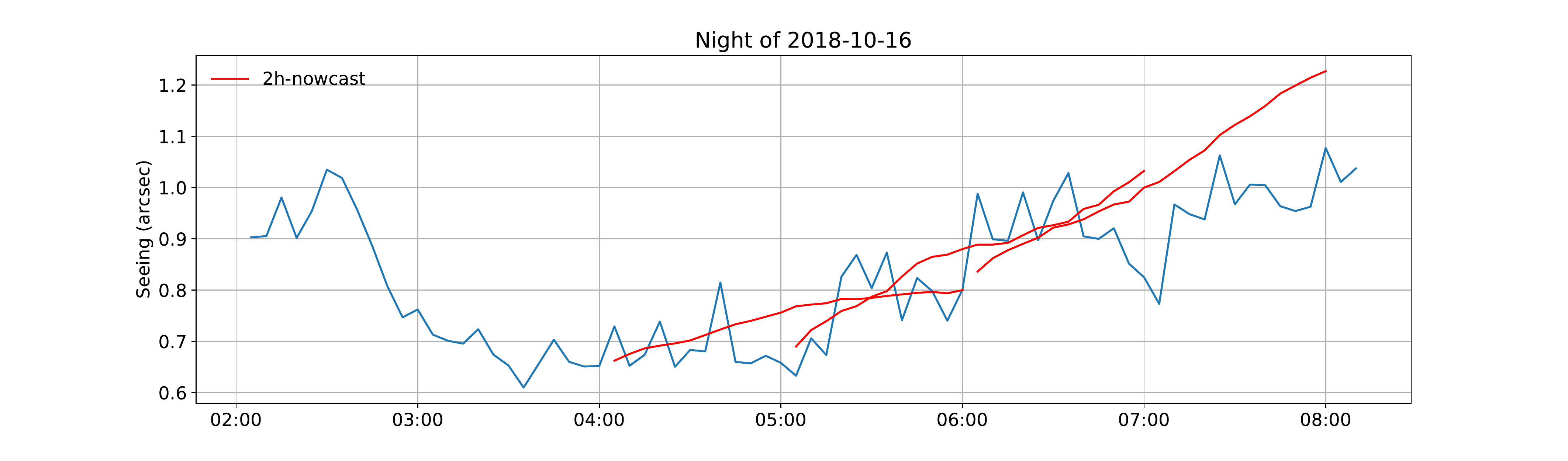}
   \includegraphics[width=0.85\hsize]{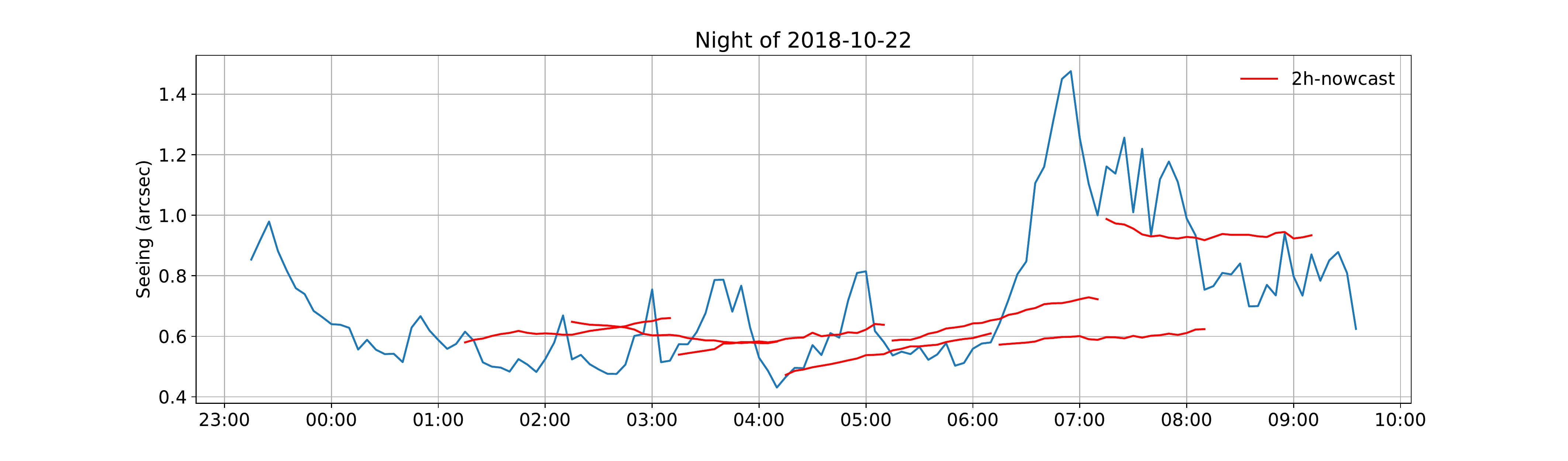}
   \includegraphics[width=0.85\hsize]{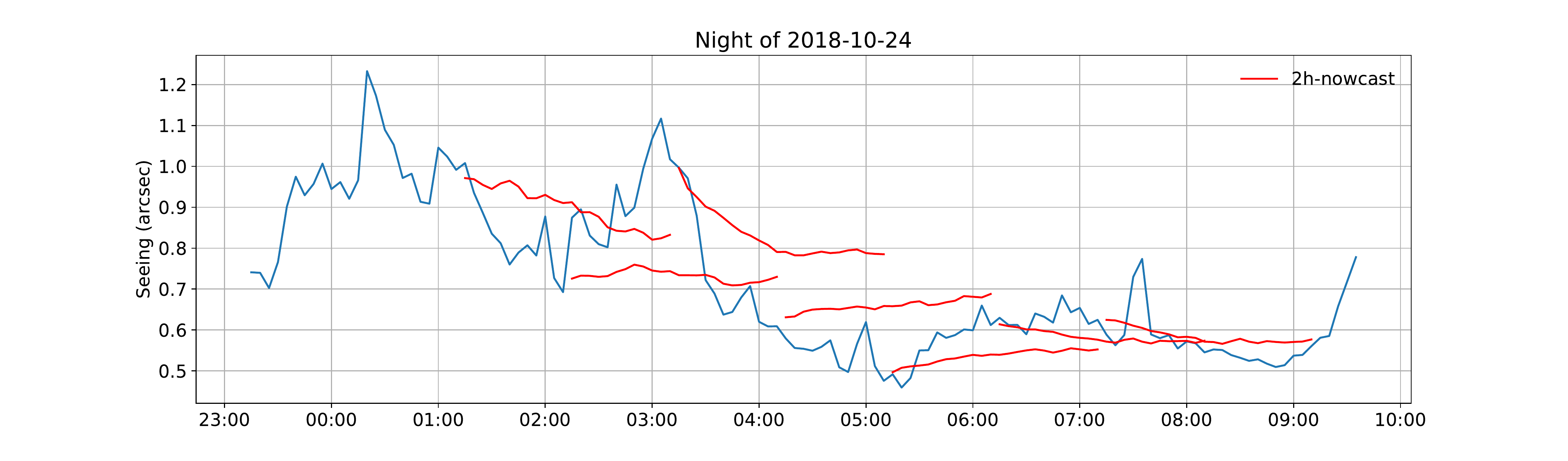}
   \includegraphics[width=0.85\hsize]{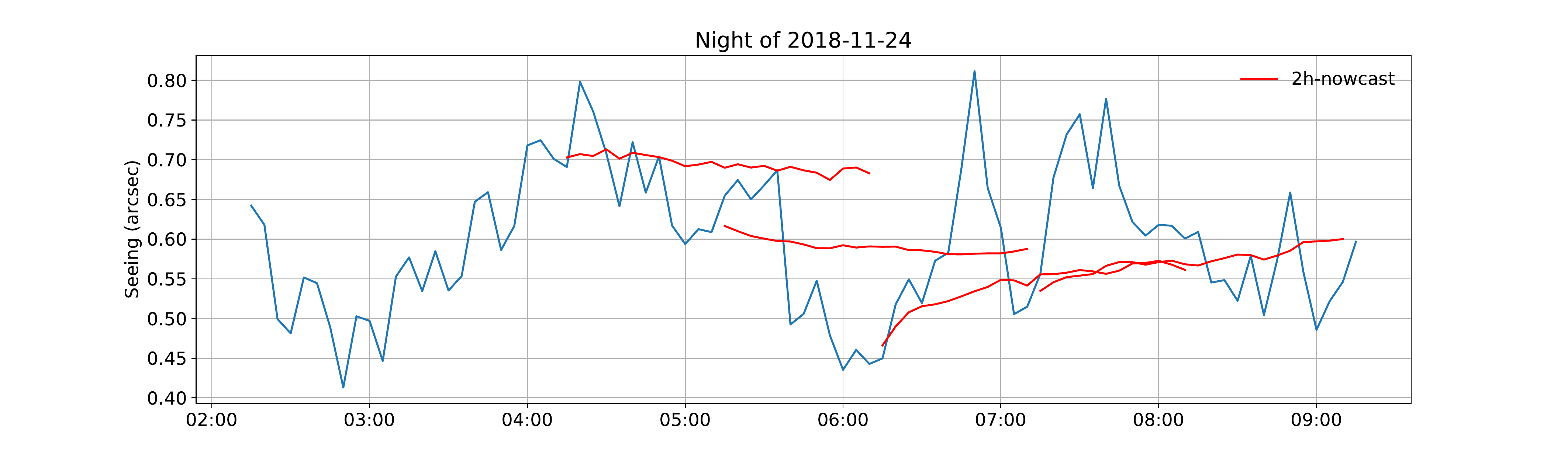}
	\end{center}
   \caption[ex] 
   { Examples of nowcasts done with the RF algorithm (red lines) for different nights from the validation data set. The blue line shows the seeing as measured by the DIMM seeing monitor. For clarity only a few nowcasts spaced by 1h have been plotted.
   \label{fig_example_RF} 
   }
   \end{figure}

We compared the accuracy of both algorithms with a very basic scenario, where we assumed that the seeing is constant in the next 2h, and the value of this constant is taken as the average seeing over the past 15min. This scenario could typically represent the behaviour of a night astronomer trying to guess the next 2h of seeing at the beginning of the night. This comparison is shown in the right plot of Fig. \ref{fig_RMSE_RF}. It shows that RF or MLP brings almost no improvement with respect to this constant scenario for a prediction below 40min, or to say it differently, that a constant scenario already achieves a good performance level compared to machine learning algorithms. For a prediction over a longer timescale, MLP brings a modest improvement. The relative improvement after 2h compared to the constant scenario is 21\% for the MLP and 10\% for the RF. 
If we look at some individual predictions, as illustrated in Fig. \ref{fig_example_RF}, we see that the high-frequency seeing variations, also called bursts in seeing, are not captured by the RF or MLP algorithms (see for instance the nights of 2018-10-22 or 2018-11-24 on the third and fourth vertical panel of Fig. \ref{fig_example_RF}). Looking at the statistical variability of the seeing confirms this statement. The variability of the seeing in a given time window is plotted in Fig. \ref{fig_seeing_variability} as a function of the length of the time window. Although the variability is low on average or even more in the sense of the median, e.g. below 0.1" over 40min, this variability exceeds 0.2" in 10\% of the cases and 0.4" in 1\% of the cases. Failing to capture those cases but providing a reasonable prediction in the other smoother cases is sufficient to explain the only modest improvement of RF and MLP over the constant scenario. Unfortunately for the observatory, those rapid increases of the seeing are also responsible for most of the failures reported in the executions of the programs and therefore represent cases where an alert would be the most valuable.

A question that naturally arises is whether with the input data presented here, the accuracy can be further improved by fine-tuning the algorithms or not. One could indeed question whether the machine-learning algorithms presented here can be further improved, because with the input data currently used, there is no harbinger that can be used to foresee a seeing burst, or whether additional training with an architecture more adapted to the nature of the problem could provide better results. We should here distinguish the case of RF and MLP. For RF, no improvement is shown using only half the data available or all of it. Likewise including additional input parameters (for instance going from 11 input variables to 13) does not yield a significant improvement, meaning that this technique likely reached its limits. This is probably due to the nature of the algorithm underlying the RF regression, which tends to build a linear extrapolation of the time-series rather than capture both low frequency trends and the high frequency stochasticity. On the other hand, the MLP results improve with larger datasets and denser networks, so there is probably plenty of room for improvements with more data and an optimal architecture. This would require a larger development time and the scope of this study was limited to investigate the accuracy of simple off-the-shelf machine learning algorithms hence our choice to focus on a dense network of fully connected layers. 

\section{Prospects and future work}

As discussed in the previous section, the current results show that there is plenty of room for improvements in terms of network architecture. In addition, the usability in operations for the observatory is another factor to consider and here again other approaches are worth considering. Last, we highlight that there is an inherent limit due to the chaotic nature of the variables to predict, the turbulence, and therefore present a possible way forward to address this issue.

\subsection{Optimising the architecture of the MLP}

Our tests suggest that the MLP shows plenty of room for improvement. Optimising the number of hidden layers of the neural network or the number of nodes in each layer still represent unexplored territories. The training set available here is finite, and therefore it might become a limiting factor in this exploration. However novel data augmentation methods can help in this respect, such as Generative Adversial Networks (GANs) or autoencoders. These algorithms also have the potential to deal better with stochasticity in the data. Recent progress has been made at predicting the behaviour of chaotic systems using reservoir computing\cite{Reservoircomputing,Reservoircomputing2}, a specific type of recurrent neural networks. One could also consider generating synthetic data from numerical weather predictions to further train the network, on the conditions that these synthetic data is close enough to real data. 

\subsection{Data sampling}

The impact of the data resampling has not been explored yet. While most input data used here are produced every minute (for atmospheric data) or 1 to 2 minutes (for the DIMM or MASS-DIMM), the results shown in Fig. \ref{fig_RMSE_RF} used data resampled every 5min mainly to ease the synchronisation of all the sources. Investigating the impact of such resampling is an important next step to make sure we do not remove important high-frequency signals that carry information on the seeing dynamics that we are trying to predict. Fig. \ref{fig_seeing_variability} (right) suggests indeed that on a timescale of 5min the seeing variability is already very large in the 10\% worst cases, with a RMS beyond $\sim0.2$". Including this seeing variability in the input parameter of the neural network is therefore an interesting and easy next step to test whether this contains information on the possible future occurrence of bursts in seeing. In this respect, it is interesting to note that a fine sampling of 2s in the sky temperature is required 
to estimate the sky transparency, as shown in \cite{Kerber2016} as lower resolution dilutes the cloudiness information.  

\subsection{A better loss function optimised for the operations of the observatory}

So far we have considered only the root mean square error (RMSE) for the loss function used in the optimisation process. This neglects however an important aspect of astronomical observations: an error in the estimation does not have the same impact for operations depending on the absolute seeing value. It has probably no consequence if the seeing is above 1.4" but can lead to time losses in any other case, especially in good conditions that the observatory cannot afford to lose.  
This problem is similar to dealing with imbalanced data\cite{He2009}, where increased accuracy is expected for very good conditions which are also less common in the training set. We see two possibilities to address this problem. The first one would be to use the same loss function as what is used in operations at the observatory. For the instance for the seeing, seven seeing categories are used in operations, from 'seeing$<0.5$"' for the stringiest to 'any conditions' for the loosest category. We could consider implementing a loss function based on those categories, that would penalise a prediction outside the range in which the observation falls. 

Alternatively, we could also take a  different approach where we compute the probability distribution that the seeing stays in a given range of values for a given time window, rather than compute the time-series directly. This present two advantages, first it absorbs part of the high frequency stochasticity into integral regions, and second it answers a more specific and useful question more operation-oriented for decision-making. We investigated such an approach in the framework of the Long Short Term Memory\cite{LSTM} (LSTM) algorithm, a specific type of recurrent neural network. An example of result is shown in Fig. \ref{fig_LSTM} (top). Note here that the seeing categories were not matched to those used in operations, this would therefore yield a much higher benefit for the operations. With this probabilistic approach, we show in Fig. \ref{fig_LSTM} (bottom) how the LSTM performs compared to the MLP. It provides a smaller prediction error for most seeing ranges.

  \begin{figure} [h]
   \begin{center}
   \includegraphics[width=0.7\hsize]{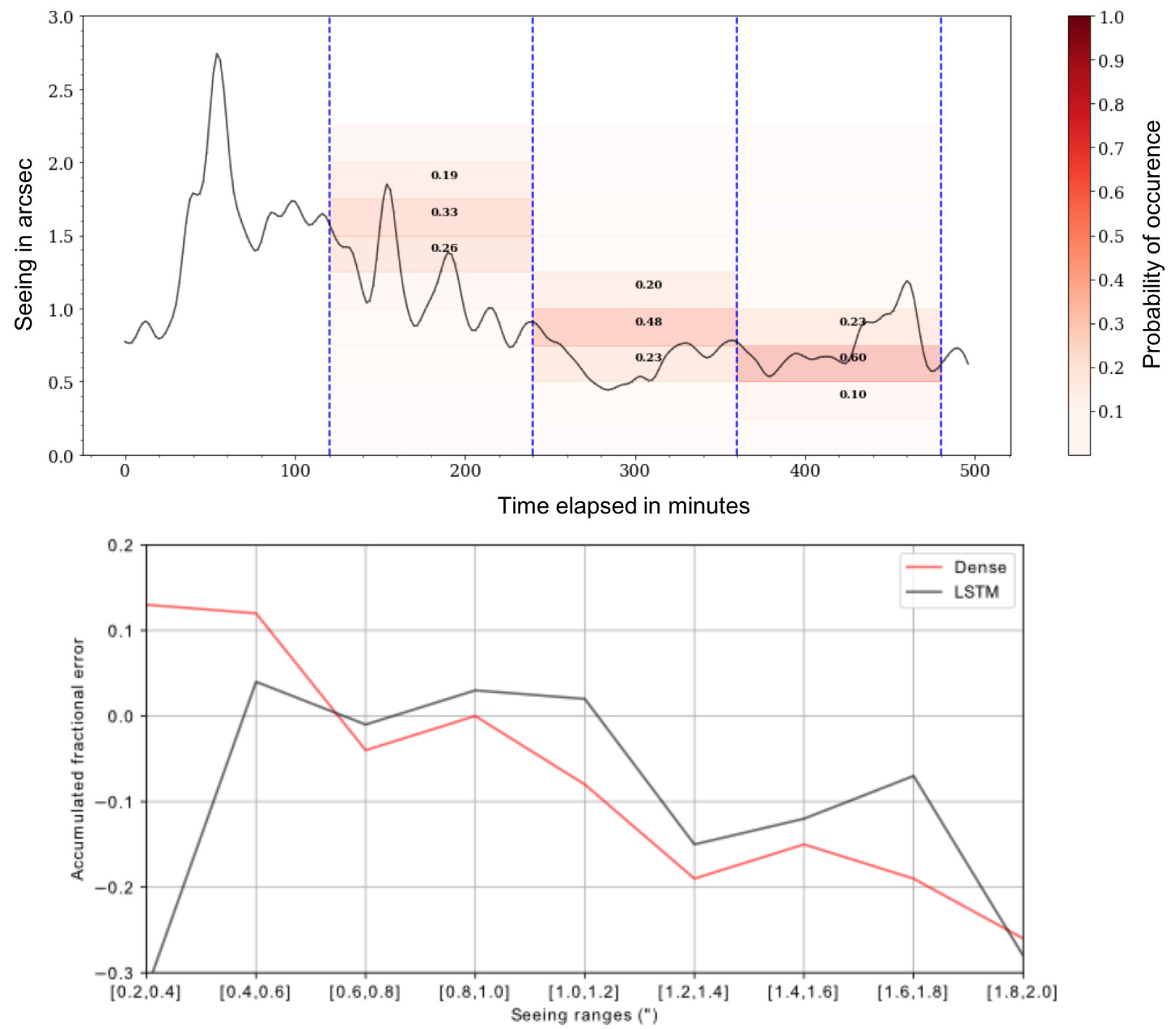}
	\end{center}
   \caption[ex] 
   { \label{fig_LSTM} 
   Example of range prediction using the LSTM algorithm\cite{LSTM}. The black line shows the seeing measurement from the DIMM. The vertical dashed blue lines delimit 3 nowcast periods of 2h each where the red colour shade indicates the likelihood of the seeing being in different ranges of 0.2" wide between 0 and 3".  \label{fig_LSTM} }
   \end{figure} 
   

\subsection{Adding spatial information: triangulation}

\begin{figure} 
   \begin{center}
   \includegraphics[width=1.\hsize]{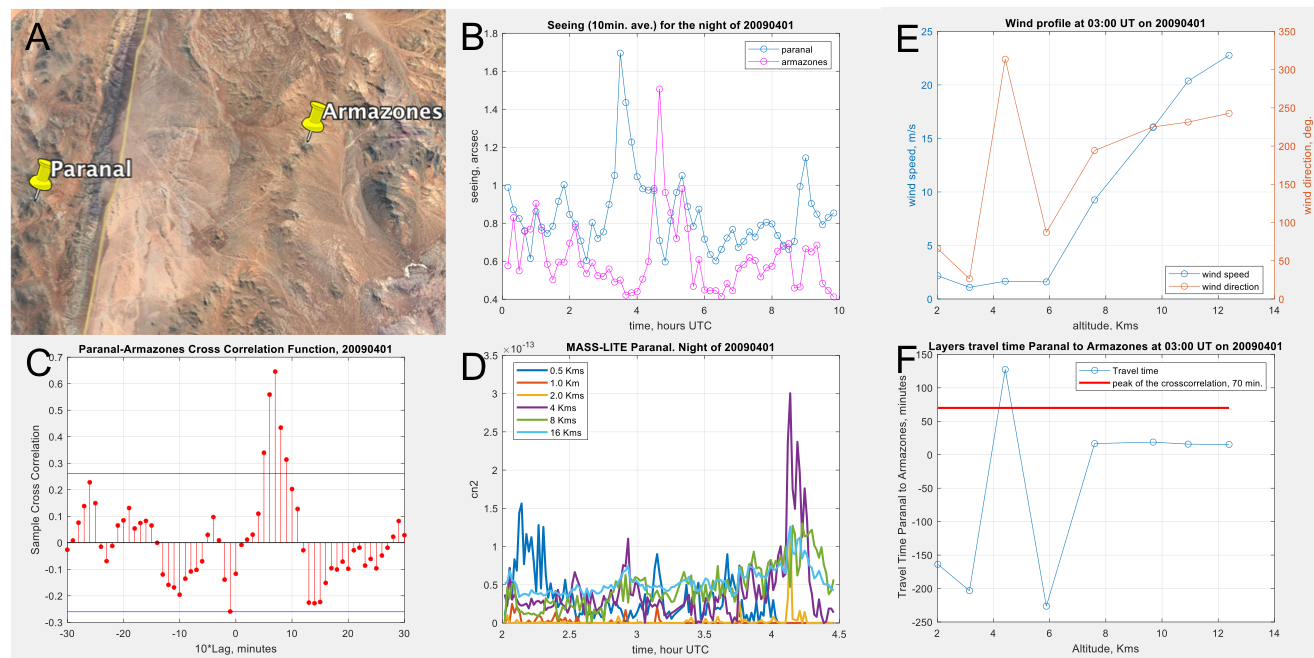}
	\end{center}
   \caption[ex] 
   {Triangulation between Paranal and Armazones. A: location of the 2 sites, 22km apart (North up). B: Seeing measured in both site on the night of October, 1 2009. C. Cross-correlation between the seeing at Paranal and Armazones, showing a peak at +70min. D: Paranal MASS measurements showing a peak of the turbulence around 4:00UT in a layer at 4km. E: Wind velocity and direction as a function of the altitude, as provided by the ECMWF forecasts. F: Expected travel time of the front from Paranal to Armazones for each layer defined in graph E as a function of the altitude (blue curve). We overplotted the 70min peak in cross-correlation as a vertical red line  \label{fig_triangulation}. Both curves cross at about 4km, compatible with the 4km turbulent layer detected y the MASS at 4:00UT (graph D).}
   \end{figure}

The reason why the accuracy of the prediction is only modest compared to a constant scenario is because the algorithm fails at capturing bursts of seeing. These bursts are likely difficult to predict because there is little or no information beforehand that can betray the imminence of their arrival. They might even come from atmospheric layers in altitude where there is no sensor to measure the changing meteorological conditions. One solution to this problem is a nowcast system with additional spatial information, ideally from a grid of weather and turbulence monitors spread around the observatory. At Cerro Paranal, a number of nearby peaks within 30km are equipped with such sensors. We will illustrate how useful these data can be in the case of a set of data gathered at Cerro Paranal and Cerro Armazones, the future location of the ELT  23km East of Cerro Paranal as shown in Fig. \ref{fig_triangulation} A, on the night of October 1 2009.  Both Paranal and Armazones seeing monitors detected a peak separated by about 70min (Fig. \ref{fig_triangulation} B and C). Ancillary data of the wind profile above Paranal (coming from ECMWF predictions) showed that such a turbulent layer was compatible with the wind speed and direction located at about 4km in altitude  (Fig. \ref{fig_triangulation} E and F). The MASS low-resolution turbulence profiler located in Paranal is indeed compatible with a  layer at 4km developing a strong turbulence around 4:00UT (Fig. \ref{fig_triangulation} D) when the seeing peak was detected in Paranal. Such an example shows that a combination of spatial (different sites) and temporal meteorological and turbulence measurements can lead to a drastic improvement in the prediction of seeing peaks.  Another example was already described in Navarette et al. 2011\cite{Navarrete2011}.

The lack of spatial information could also be compensated by the use of numerical weather predictions. For instance, the surface layer parameters are known to be difficult to predict due to the need for high spatial and temporal resolution to describe the phenomena close to the ground and due to the scarcity of initialisation parameters. They could be measured by dedicated  sensors on the ground. The free-atmosphere parameters, on the other hand, could be predicted by numerical weather prediction codes, since they are more accurately forecasted than surface layer parameters. Then an artificial intelligence algorithm could combine those two input sources to nowcast the turbulence. 

\section{Conclusions} 

We presented here the motivations for a system of turbulence nowcast for the Paranal observatory.  We investigated the accuracy of a purely data-driven system, making use of three different off-the-shelf machine learning algorithms: random forests, multilayer perceptrons and long-short term memory. We showed that such algorithms can provide a nowcast over 2h, with a modest accuracy improvement with respect to a constant prediction scenario. We showed that the main reason for this modest performance is the high-frequency behaviour of the turbulence, with bursts in seeing, that are not captured by these basic algorithms. There exists much room for improvement in fine-tuning the architecture of the neural networks to better adapt to the chaotic nature of the turbulence. In addition, we presented an interesting technique combining spatial information from other nearby sites to enable the nowcast of those seeing bursts.

\acknowledgments 
 
J.M. thanks ESO staff and technical operators at Paranal Observatory, especially the Software team to allow to use the ESO Data Lab\cite{Pena2018} for the analysis presented in this paper. He thanks P. Figueira for interesting discussions on time losses due to the seeing. This research made use of the following python packages: matplotlib, pandas, scikit-learn\cite{scikit-learn} and keras\cite{Keras}. 

\bibliography{biblio_ao4elt} 

\newcommand{\noop}[1]{}
\begin{thebibliography}{10}

\bibitem{Patat2017}
{Patat}, F., {Hussain}, G., et~al., ``{Period 100: The Past, Present and Future
  of ESO Observing Programmes},'' {\em The Messenger}~{\bf 169},  5--10 (Sep
  2017).

\bibitem{Woillez2019}
{Woillez}, J., {Abad}, J.~A., et~al., ``{NAOMI: the adaptive optics system of
  the Auxiliary Telescopes of the VLTI},'' {\em \aap}~{\bf 629},  A41 (Sep
  2019).

\bibitem{Woillez2016}
{Woillez}, J., {Alonso}, J., et~al., ``{The 2nd generation VLTI path to
  performance},'' in [{\em \procspie}{\nolinebreak\hspace{0.1em}]},  {\em
  Society of Photo-Optical Instrumentation Engineers (SPIE) Conference Series}
  {\bf 9907},  990706 (Aug 2016).

\bibitem{Madec2018}
{Madec}, P.~Y., {Arsenault}, R., et~al., ``{Adaptive Optics Facility: from an
  amazing present to a brilliant future...},'' in [{\em
  \procspie}{\nolinebreak\hspace{0.1em}]},  {\em Society of Photo-Optical
  Instrumentation Engineers (SPIE) Conference Series} {\bf 10703},  1070302
  (Jul 2018).

\bibitem{Milli2017_SPHERE}
{Milli}, J., {Mouillet}, D., et~al., ``{Performance of the extreme-AO
  instrument VLT/SPHERE and dependence on the atmospheric conditions},'' {\em
  ArXiv e-prints}  (Oct. 2017).

\bibitem{Rejkuba2018}
{Rejkuba}, M., {Tacconi-Garman}, L.~E., et~al., ``{Should I stay, or should I
  go? Service and Visitor Mode at ESO's Paranal Observatory},'' {\em The
  Messenger}~{\bf 173},  2--6 (Sep 2018).

\bibitem{Bonnet2018}
{Bonnet}, H., {Biancat-Marchet}, F., et~al., ``{Adaptive optics at the ESO
  ELT},'' in [{\em \procspie}{\nolinebreak\hspace{0.1em}]},  {\em Society of
  Photo-Optical Instrumentation Engineers (SPIE) Conference Series} {\bf
  10703},  1070310 (Jul 2018).

\bibitem{Veilleit2016}
{Veillet}, C., {Ashby}, D.~S., et~al., ``{LBTO's long march to full operation:
  step 2},'' in [{\em \procspie}{\nolinebreak\hspace{0.1em}]},  {\em Society of
  Photo-Optical Instrumentation Engineers (SPIE) Conference Series} {\bf 9910},
   99100S (Aug 2016).

\bibitem{Turchi2017}
{Turchi}, A., {Masciadri}, E., and {Fini}, L., ``{Forecasting surface-layer
  atmospheric parameters at the Large Binocular Telescope site},'' {\em
  \mnras}~{\bf 466},  1925--1943 (Apr 2017).

\bibitem{Lafore1998}
{Lafore}, J.~P., {Stein}, J., et~al., ``{The Meso-NH Atmospheric Simulation
  System. Part I: adiabatic formulation and control simulations},'' {\em
  Annales Geophysicae}~{\bf 16},  90--109 (Jan 1998).

\bibitem{Masciadri1999}
{Masciadri}, E., {Vernin}, J., and {Bougeault}, P., ``{3D mapping of optical
  turbulence using an atmospheric numerical model. I. A useful tool for the
  ground-based astronomy},'' {\em \aaps}~{\bf 137},  185--202 (May 1999).

\bibitem{Masciadri2013}
{Masciadri}, E., {Lascaux}, F., and {Fini}, L., ``{MOSE: operational forecast
  of the optical turbulence and atmospheric parameters at European Southern
  Observatory ground-based sites - I. Overview and vertical stratification of
  atmospheric parameters at 0-20 km},'' {\em \mnras}~{\bf 436},  1968--1985
  (Dec 2013).

\bibitem{Lascaux2013}
{Lascaux}, F., {Masciadri}, E., and {Fini}, L., ``{MOSE: operational forecast
  of the optical turbulence and atmospheric parameters at European Southern
  Observatory ground-based sites - II. Atmospheric parameters in the surface
  layer 0-30 m},'' {\em \mnras}~{\bf 436},  3147--3166 (Dec 2013).

\bibitem{Lascaux2015}
{Lascaux}, F., {Masciadri}, E., and {Fini}, L., ``{Forecast of surface layer
  meteorological parameters at Cerro Paranal with a mesoscale atmospherical
  model},'' {\em \mnras}~{\bf 449},  1664--1678 (May 2015).

\bibitem{Turchi2019}
{Turchi}, A., {Masciadri}, E., et~al., ``{Forecasting water vapour above the
  sites of ESO's Very Large Telescope (VLT) and the Large Binocular Telescope
  (LBT)},'' {\em \mnras}~{\bf 482},  206--218 (Jan 2019).

\bibitem{Masciadri2017}
{Masciadri}, E., {Lascaux}, F., et~al., ``{Optical turbulence forecast: ready
  for an operational application},'' {\em \mnras}~{\bf 466},  520--539 (Apr
  2017).

\bibitem{Osborn2018_GSM}
{Osborn}, J. and {Sarazin}, M., ``{Atmospheric turbulence forecasting with a
  general circulation model for Cerro Paranal},'' {\em \mnras}~{\bf 480},
  1278--1299 (Oct. 2018).

\bibitem{Sarazin1990}
{Sarazin}, M. and {Roddier}, F., ``{The ESO differential image motion
  monitor},'' {\em \aap}~{\bf 227},  294--300 (Jan. 1990).

\bibitem{Kornilov2007}
{Kornilov}, V., {Tokovinin}, A., et~al., ``{Combined MASS-DIMM instruments for
  atmospheric turbulence studies},'' {\em \mnras}~{\bf 382},  1268--1278 (Dec.
  2007).

\bibitem{Kerber2012}
{Kerber}, F., {Rose}, T., et~al., ``{A water vapour monitor at Paranal
  Observatory},'' in [{\em \procspie}{\nolinebreak\hspace{0.1em}]},  {\em
  Society of Photo-Optical Instrumentation Engineers (SPIE) Conference Series}
  {\bf 8446},  84463N (Sep 2012).

\bibitem{Roddier1981}
{Roddier}, F., ``{The effects of atmospheric turbulence in optical
  astronomy},'' {\em Progress in optics.~Volume 19.~Amsterdam, North-Holland
  Publishing Co., 1981, p.~281-376.}~{\bf 19},  281--376 (1981).

\bibitem{Masciadri2001}
{Masciadri}, E. and {Garfias}, T., ``{Wavefront coherence time seasonal
  variability and forecasting at the San Pedro M{\'a}rtir site},'' {\em
  \aap}~{\bf 366},  708--716 (Feb 2001).

\bibitem{Sarazin2002}
{Sarazin}, M. and {Tokovinin}, A., ``{The Statistics of Isoplanatic Angle and
  Adaptive Optics Time Constant derived from DIMM Data},'' in [{\em European
  Southern Observatory Conference and Workshop
  Proceedings}{\nolinebreak\hspace{0.1em}]},  {Vernet}, E., {Ragazzoni}, R.,
  et~al., eds., {\em European Southern Observatory Conference and Workshop
  Proceedings} {\bf 58},  321 (2002).

\bibitem{Breiman1984}
Breiman, L., Friedman, J., et~al.,  [{\em Classification and Regression
  Trees}{\nolinebreak\hspace{0.1em}]}, The Wadsworth and Brooks-Cole
  statistics-probability series, Taylor \& Francis (1984).

\bibitem{scikit-learn}
Pedregosa, F., Varoquaux, G., et~al., ``Scikit-learn: Machine learning in
  {P}ython,'' {\em Journal of Machine Learning Research}~{\bf 12},  2825--2830
  (2011).

\bibitem{Keras}
Chollet, F. et~al., ``Keras.'' \url{https://keras.io} (2015).

\bibitem{Kingma2014}
{Kingma}, D.~P. and {Ba}, J., ``{Adam: A Method for Stochastic Optimization},''
  {\em arXiv e-prints} ,  arXiv:1412.6980 (Dec 2014).

\bibitem{Reservoircomputing}
Pathak, J., Hunt, B., et~al., ``Model-free prediction of large spatiotemporally
  chaotic systems from data: A reservoir computing approach,'' {\em Phys. Rev.
  Lett.}~{\bf 120},  024102 (Jan 2018).

\bibitem{Reservoircomputing2}
{Chattopadhyay}, A., {Hassanzadeh}, P., et~al., ``{Data-driven prediction of a
  multi-scale Lorenz 96 chaotic system using a hierarchy of deep learning
  methods: Reservoir computing, ANN, and RNN-LSTM},'' {\em arXiv e-prints} ,
  arXiv:1906.08829 (Jun 2019).

\bibitem{Kerber2016}
{Kerber}, F., {Querel}, R.~R., et~al., ``{Through thick and thin: quantitative
  classification of photometric observing conditions on Paranal},'' in [{\em
  \procspie}{\nolinebreak\hspace{0.1em}]},  {\em Society of Photo-Optical
  Instrumentation Engineers (SPIE) Conference Series} {\bf 9910},  99101S (Jul
  2016).

\bibitem{He2009}
He, H. and Garcia, E.~A., ``Learning from imbalanced data,'' {\em IEEE Trans.
  on Knowl. and Data Eng.}~{\bf 21},  1263--1284 (Sept. 2009).

\bibitem{LSTM}
Hochreiter, S. and Schmidhuber, J., ``Long short-term memory,'' {\em Neural
  Comput.}~{\bf 9},  1735--1780 (Nov. 1997).

\bibitem{Navarrete2011}
{Navarrete}, J., ``{The VLT dealing with the Atmosphere, a Night Operation
  point of view},'' {\em arXiv e-prints} ,  arXiv:1101.2341 (Jan 2011).

\bibitem{Pena2018}
{Pe\~na}, E., {Schmutzer}, R., et~al., ``{Framework to use modern Big Data
  Software Tools to improve operations at the Paranal Observatory},'' in [{\em
  Modeling, Systems Engineering, and Project Management for Astronomy
  VIII}{\nolinebreak\hspace{0.1em}]},  {\em \procspie} (2018).

\end{thebibliography}
\bibliographystyle{spiebib_first_3_authors} 


\end{document}